# Probing Band-Tail States in Silicon MOS Heterostructures with Electron Spin Resonance

R. M. Jock[1], S. Shankar[1,3], A. M. Tyryshkin[1], Jianhua He[1], K. Eng[2,4], K.D. Childs[2], L. A. Tracy[2], M. P. Lilly[2], M. S. Carroll[2], S. A. Lyon[1]

[1]*Department of Electrical Engineering, Princeton University*
[2]*Sandia National Laboratories*
[3]*Now at Applied Physics Department, Yale University*
[4]*Now at HRL Laboratories*

We present an electron spin resonance (ESR) approach to characterize shallow electron trapping in band-tail states at Si/SiO$_2$ interfaces in metal-oxide-semiconductor (MOS) devices and demonstrate it on two MOS devices fabricated at different laboratories. Despite displaying similar low temperature (4.2 K) peak mobilities, our ESR data reveal a significant difference in the Si/SiO$_2$ interface quality of these two devices, specifically an order of magnitude difference in the number of shallow trapped charges at the Si/SiO$_2$ interfaces. Thus, our ESR method allows a quantitative evaluation of the Si/SiO$_2$ interface quality at low electron densities, where conventional mobility measurements are not possible.

Understanding the effect of the Si/SiO$_2$ interface on the electronic properties of two-dimensional (2D) electrons in metal-oxide-semiconductor (MOS) heterostructures has been a long-standing issue in MOS physics.[1-4] Recently this issue has taken on a new importance due to the interest in single or few electron quantum devices[5-7]. Imperfections near the Si/SiO$_2$ interface, such as trapped charges and interface roughness, lead to potential fluctuations,[8,9] that can confine electrons at local potential minima. These shallow confined electron states, having energies of a few meV,[10,11] can severely limit the control of few electron devices through electrostatic gating as needed to manipulate electron charge and spin states. Therefore, the interface quality of the heterostructure needs to be assessed and optimized. Typically, electron mobility is used as a figure of merit to characterize interface quality. Mobility measurements, however, are performed in the presence of many electrons, whereas in a few electron quantum devices the interface will be largely depleted of electrons. Therefore, other methods must be developed to allow a complete evaluation of the interface quality at low electron densities where few electron quantum devices will operate.

In this work we use electron spin resonance (ESR) to directly probe the localized states below the band edge, determining the density of states within a few meV of the conduction band edge in MOS devices and thus characterizing interface quality. We compare two silicon MOS field-effect-transistors (MOSFET) with similar low temperature (4.2 K) peak mobilities. One device was fabricated in-house (in a university clean room) and ESR measurements on this device have been previously reported for temperatures between 2 and 10 K.[11] These measurements are now extended to lower temperatures (370mK) and are compared to measurements of a second device made in a silicon foundry operated by Sandia National Laboratories. We find that despite showing similar peak mobilities, the density and depth of shallow confined electron states differ by an order of magnitude in these two devices.

Sample A (fabricated at Princeton) is an n-channel accumulation MOSFET fabricated on an isotopically-enriched 25 μm $^{28}$Si (001) epi-wafer (Isonics, residual 800 ppm of $^{29}$Si, background doping of $10^{14}$ phosphorous per cm$^3$). The device has phosphorus implanted source-drain contacts, a 110 nm dry thermal gate oxide, and a 100 nm Ti/Au metal gate. The device was annealed for 15 min at 1050°C in N$_2$ after oxidation, and received a post metallization anneal for 25 min at 450°C in forming gas. The MOSFET's gate area is large (0.4 x 2 cm$^2$) in order to obtain a sufficient ESR signal from the 2D electrons. Transport measurements show a threshold voltage of 1 V and a peak Hall mobility of 14,000 cm$^2$V$^{-1}$s$^{-1}$ at 4.2 K. Sample B (fabricated at Sandia) is an n-channel inversion MOSFET made on a 5000 Ohm-cm p-type (001) natural silicon wafer (Topsil). The device has arsenic implanted source-drain contacts, a 35 nm dry thermal gate oxide, and a large area (0.2 x 2 cm$^2$) 200 nm n+ poly-silicon gate coated with tungsten. Anneals were performed after gate oxidation for 30 min at 900°C in N$_2$, following the poly-silicon deposition for 13 min in O$_2$ and 30 min in N$_2$ at 900°C, and after metallization for 30 min in forming gas then 30 min in N$_2$ at 400°C. Transport measurements give a threshold voltage of 0.3 V and peak Hall mobility of 10,000 cm$^2$V$^{-1}$s$^{-1}$ at 4.2 K. Both devices were fabricated with an extended length in order to keep the metal contacts away from the resonator.[12]

Continuous wave ESR measurements were performed using a Bruker Elexsys580 spectrometer operating at X-band frequency (approximately 9.6 GHz). A $^3$He cryostat (Janis Research) was used to maintain sample temperatures in the range between 370 mK and 3 K. ESR measurements of sample B display a gate-dependent signal having a g-factor of 2.0001 and a line-width of 0.6 G, similar to that reported previously in sample A (g-factor of 1.9999 and line-width of 0.2 G) for 2D electrons at Si/SiO$_2$ interfaces.[10,11]

Figure 1 (insert) illustrates a typical dependence of the number of unpaired 2D electron spins (calculated as the integrated ESR signal intensity) as a function of the applied gate voltage ($V_G$) in sample B at 3 K, covering a broad range of $V_G$ both below and above threshold. The dependence is similar to that reported before for sample A at comparable temperatures.[11] For $V_G$ above threshold, or equivalently the Fermi energy of 2D electrons ($E_F$) above the conduction band edge ($E_C$), the density of states (DOS) of 2D electrons is known to be constant and energy independent. Therefore as $V_G$ increases above threshold ($E_F$ increases), the number of mobile electrons increases, but the number of unpaired electron spins, lying within $g\mu_B B_0$ of the Fermi surface, remains constant as observed in our experiment (Figure 1, insert). Here, g is the electron g-factor, $\mu_B$ is the Bohr magneton, and $B_0$ is the applied magnetic field. For $V_G$ below threshold ($E_F < E_C$), when $E_F$ moves into the localized states below the band edge, two different regimes should be resolved. When $V_G$ is reduced slightly below threshold, the trapped electrons at the Fermi surface can still hop ($k_BT > E_C - E_F$) and escape to the source-drain contacts, allowing the trapped electrons at the interface to equilibrate with the contacts. Figure 1 (insert) shows a monotonic decrease in the number of confined unpaired spins as $V_G$ decreases from threshold to 50 mV at 3 K, as expected. The situation changes when $V_G$ is well below threshold and the localized electrons at the Fermi surface no longer have sufficient thermal energy to escape from the confining potentials ($k_BT \ll E_C - E_F$). At this point the Fermi energy, $E_F$, becomes pinned and the electrons are frozen into localized states. As $V_G$ continues to decrease, the number of trapped electrons, and therefore unpaired spins, remains constant and the signal intensity becomes independent of gate voltage.

Illuminating the device with above band gap light creates holes, which neutralize electrons confined in the localized states. Figure 1 (main) illustrates that the number of trapped spins decreases upon illumination (1050 nm) when measured at low gate voltages for sample B at 1 K and 3 K. In sample B the gate voltage must be made more negative by ~30meV to reduce the illuminated ESR signal to zero. By comparing the dark and post-illuminated curves, the voltage at which the two plots diverge, or where $E_F$ becomes pinned when measured in the dark, can be found. This voltage, which we define as V*, is the gate voltage at which the Fermi level is aligned with the energy of the shallowest states in the band-tail that are unable to thermally de-trap ($k_BT \sim E_C - E_F$). The characteristic gate voltages, V*, at which the Fermi energy is pinned were measured for both samples A and B over a range of temperatures. At each temperature the extracted V* was used to estimate the density of trapped electrons per unit area using the known capacitances of the devices (e.g. $e \cdot n^* = C_{ox}V^*$, where gate capacitances were obtained from Hall measurements above threshold). Note that using this method the derived electron densities are measured as a charge, rather than through spin counting, and are thus independent of any spin pairing. The measured n* for both devices are plotted against temperature in Figure 2. We notice that the density, and thus V*, increases approximately as 1/T. At lower temperatures the Fermi level is pinned at a higher energy and therefore a larger number of electrons is confined. Comparing the two devices, we notice that sample A displays a significantly higher density of confined electrons.

The data in Figure 2 can be used to calculate the density of confined states below the conduction band edge in both devices. To do so, we first need to convert the temperature scale (x-axis) in Figure 2 into the electron confinement energy scale. We use a simplified Shockley-Read-Hall model and express the escape time, τ, for confined electrons as a function of $E_F$ with respect to the conduction band and temperature, T.[11,13] The equation is then inverted, to determine how $E_F$ changes with T, assuming τ = $10^4$ seconds (longer than our measurement time). The numerical solution of this problem gives an approximate dependence $E_C - E_F = 10 \cdot k_BT$ for the confinement depth of the shallowest states occupied at each given temperature. After converting the temperature to energy scale, a simple numerical differentiation of the data in Figure 2 gives the DOS below the conduction band edge. The results for both devices are depicted in Figure 3 along with the calculated DOS for mobile 2D electrons at the band edge at a (100) Si/SiO$_2$ interface. We note that our results agree with the expected value at the band edge. The samples behave similarly near the conduction band, with sample B displaying a somewhat steeper slope. However, at confinement energies deeper than 2.25 meV sample B has no detectable confined electrons states, whereas sample A displays a wide range of confinement energies with a significant density of localized states even at a 10 meV depth.

Further information about the nature of these shallow electron states is provided by the temperature behavior of the ESR signal. Previous measurements on sample A have shown a Pauli-like susceptibility with the MOSFET biased above threshold.[10] However, when biased below threshold the spin susceptibility displays a clear 1/T dependence, indicating a Curie-like behavior. Our recent measurements indicate that this persists down to 370 mK and is evident in sample B as well. Since Curie law susceptibility is a characteristic of isolated, independent electrons, this result suggests a singly confined nature of the electron states.

Knowing the density and confinement depth of trapped electrons at the interface, an estimate can be made for the average distance between and the upper bound in size of the trapped electron wavefunctions in each device. Considering the electrons with confinement stronger than 1 meV, densities of 1.25 x $10^{11}$ cm$^{-2}$ and 4.0 x $10^{10}$ cm$^{-2}$ can be estimated for samples A and B, respectively. These densities correspond to the average distance between electrons of 28 nm in sample A and 50 nm in sample B. Taking a very simple picture of parabolic potential minima (a common approximation for lateral quantum dots[14]) crossing at the confinement energy and separated by the average spacing,[15] we can then estimate an upper bound for electron confinement dimensions of 28 nm and 37 nm in samples A and B, respectively. If we consider deeper traps the difference between these two devices will be greater; the maximum permissible trap depth will depend on the geometry and fields used for controlling a particular single electron device.

In conclusion, we have used electron spin resonance to measure the number of trapped electrons and deduce the density of states below the conduction band edge in silicon MOSFETs. Our results demonstrate the ESR method as a valuable tool for probing 2D band-tail states and for evaluating the interface quality in FET structures that is of special importance when building a few electron devices. Comparing two MOSFETs fabricated with different processes, but which appear similar from mobility measurements, we find that one of the devices displays a smaller band-tail density of confined states and a shallower characteristic confinement. This reveals a difference in device quality, which is not apparent from mobility measurements.

Work at Princeton was supported by NSA/LPS and the ARO (W911NF-04-1-0389) and by NSF through the Princeton MRSEC (DMR-0819860). Part of this work was performed at the Center for Integrated Nanotechnologies, a U.S. Department of Energy, Office of Basic Energy Sciences user facility. Sandia is a multiprogram laboratory operated by Sandia Corporation, a Lockheed Martin Co., for the United States Department of Energy under Contract No. DE-AC04-94AL85000.

[1] T. Ando, A. B. Fowler, and F. Stern, Rev. Mod. Phys. **54**, 437 (1982).
[2] Y. C. Chen, *Progress in Surface Science*, **8**, (1977).
[3] S. M. Goodnick, D. K. Ferry, C. W. Wilmsen, Z. Liliental, D. Fathy, and O. L. Krivanek, Phys. Rev. B **32**, 8171 (1985).
[4] S. Yamakawa, H. Ueno, K. Taniguchi, C. Hamaguchi, K. Miyatsuji, K. Masaki, and U. Ravaioli, J. Appl. Phys. **79**, 911 (1996).
[5] E. P. Nordberg, G. A. Ten Eyck, H. L. Stalford, R. P. Muller, R. W. Young, K. Eng, L. A. Tracy, K. D. Childs, J. R. Wendt, R. K. Grubbs, J. Stevens, M. P. Lilly, M. A. Eriksson, and M. S. Carroll, Phys. Rev. B **80**, 115331 (2009).
[6] Martin Fuechsle, S. Mahapatra, F. A. Zwanenburg, Mark Friesen, M. A. Eriksson, Michelle Y. Simmons, Nature Nanotechnology, (2010).
[7] M. Xiao, M. G. House, and H. W. Jiang, Phys. Rev. Lett. **104**, 096801 (2010).
[8] J. A. Nixon and J. H. Davies, Phys. Rev. B **41**, 7929 (1990).
[9] J. R. Brews, J. Appl. Phys. **43**, 2306 (1972).
[10] S. Shankar, A. M. Tyryshkin, S. Avasthi, and S. A. Lyon, Physica E **40**, 1659 (2008).
[11] S. Shankar, A. M. Tyryshkin, Jianhua He, and S. A. Lyon Phys. Rev. B **82**, 195323 (2010).
[12] C. Boehme and K. Lips, Physica B **376-377**, 930 (2006).
[13] W. Shockley and W. T. Read, Phys. Rev. 87, 835 (1952).
[14] Stephanie M. Reimann and Matti Manninen, Rev. Mod. Phys. **74**, 1283 (2002).
[15] S. Shankar, Electron spin coherence in bulk silicon and silicon heterostructures, Ph.D. thesis, Princeton University (2010).

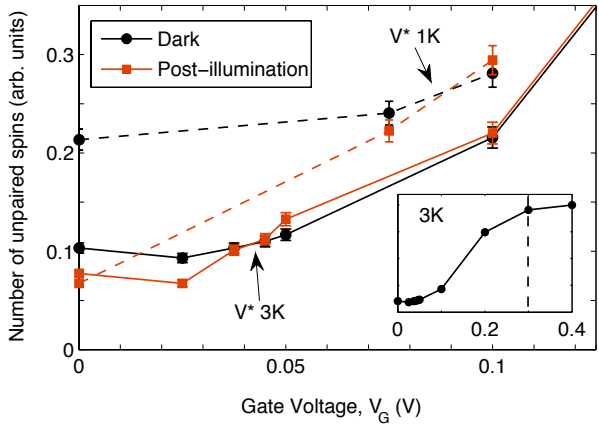

Fig 1. Number of unpaired spins as a function of gate voltage for sample B as measured at low gate voltages at 1 K (connected by dashed lines) and 3 K (connected by solid lines). At both temperatures the black curves were measured in the dark and the red curves were after above band gap illumination. Inset shows the gate voltage dependence measured at 3 K in a broad gate voltage range, including above and below threshold voltages. The threshold voltage in sample B is indicated by the vertical dashed line.

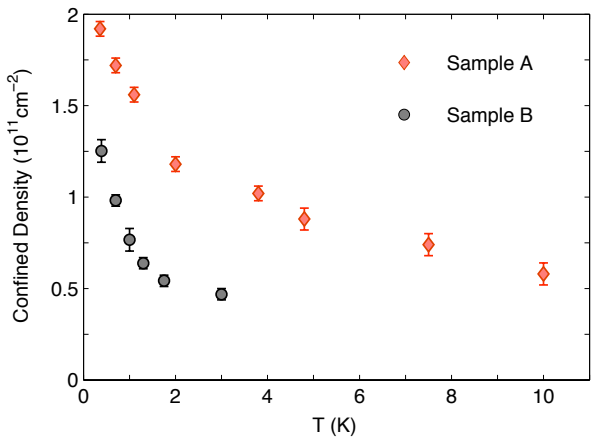

Fig 2. Density of confined electrons for samples A (red diamonds) and B (black circles) as a function of temperature.

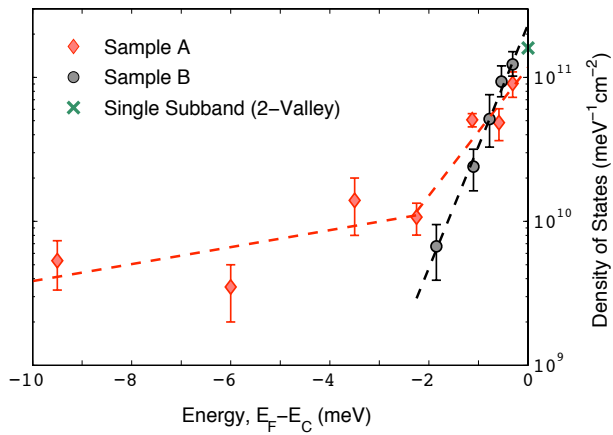

Fig 3. Density of confined electron states as a function of energy below the conduction band for samples A (red diamonds) and B (black circles). The green cross indicates the calculated value ($1.6 \times 10^{11}$ meV$^{-1}$ cm$^{-2}$) of the density of states in the conduction band for 2D electrons at a Si(100) surface (two-fold valley degeneracy). The dashed lines are linear fits in the regions below and above 2.25 meV.